\documentclass[10pt,twocolumn]{article}
\usepackage[letterpaper,left=0.72in,right=0.72in,top=0.68in,bottom=0.72in,columnsep=0.27in]{geometry}
\usepackage[T1]{fontenc}
\usepackage{newtxtext,newtxmath}
\usepackage{microtype}
\usepackage{graphicx}
\usepackage{booktabs}
\usepackage{amsmath}
\usepackage{seqsplit}
\usepackage{tikz}
\usetikzlibrary{arrows.meta,fit,positioning,shapes.geometric}
\usepackage{pgfplots}
\usepgfplotslibrary{groupplots}
\pgfplotsset{compat=1.18}
\usepackage{balance}
\usepackage[hidelinks]{hyperref}
\makeatletter
\renewcommand\section{\@startsection{section}{1}{\z@}%
  {1.1ex \@plus .25ex}{.45ex}{\normalfont\fontsize{14}{16}\bfseries}}
\renewcommand\subsection{\@startsection{subsection}{2}{\z@}%
  {.9ex \@plus .2ex}{.35ex}{\normalfont\fontsize{12}{14}\bfseries}}
\makeatother
\begin{document}
\fontsize{10}{11.8}\selectfont
\twocolumn[
\begin{center}
{\fontsize{17.3}{19.5}\selectfont Withholding the Completing Chunk: Exact Release-\allowbreak{}Boundary Equivalence for Production Streaming Guardrails\par}
\vspace{6pt}
{\large Christopher M. Frost}\\\vspace{2pt}
HEOSSI (Pte.) Ltd., Singapore\\\vspace{2pt}
\href{mailto:christopher@heossi.com}{christopher@heossi.com}\quad
\href{https://orcid.org/0009-0002-1027-1149}{ORCID 0009-0002-1027-1149}
\end{center}
\vspace{2pt}
\noindent\textbf{Abstract---} Streaming language-\allowbreak{}model output creates an enforcement boundary: a control that detects a prohibited pattern after releasing its completing chunk cannot recall it. We study a production policy in which each ordered family is the conjunction of two regular-\allowbreak{}language predicates. Incremental matching is classical. The problem is exact composition at release time across arbitrary chunk partitions, including end-\allowbreak{}of-\allowbreak{}prefix word boundaries that can change on extension. We define an ASCII-\allowbreak{}explicit policy grammar, compile each predicate to a persistent nondeterministic finite automaton (NFA), distinguish stable from provisional assertion state, apply document-\allowbreak{}order family priority, and check the decision before releasing each chunk. We show that the resulting monitor is release-\allowbreak{}boundary equivalent to an absorbing cumulative oracle for every policy in the declared grammar. Production Python and TypeScript implementations were evaluated on 101,653 partitioned cases; a public surrogate added 100,345 cases. Both campaigns produced zero oracle, cross-\allowbreak{}runtime, or intended-\allowbreak{}family mismatches. In a frozen neutral-\allowbreak{}output profile, the memoized incremental and native-\allowbreak{}regex cumulative slopes at 64-\allowbreak{}character chunks were 0.973 and 1.976. At 16,384 characters the incremental median was 30.2 ms versus 96.6 ms for native cumulative scanning at that chunk size. Native regex remained faster at 512-\allowbreak{}character chunks (12.4 versus 29.4 ms), exposing the constant-\allowbreak{}factor crossover rather than hiding it. A shared per-\allowbreak{}stream cache cap and 129-\allowbreak{}symbol alphabet bound optimization state; the campaign peaked at 364 of 4,096 without bypass. The result is policy conformance for a deterministic backstop, not evidence of semantic safety or policy completeness.

\vspace{4pt}
\vspace{9pt}
]

\section{Introduction}

An output guard has two distinct responsibilities: decide whether generated text violates its policy, and intervene before violating text crosses the release boundary. Complete-\allowbreak{}response moderation addresses the first question only after streaming text may already have escaped. Learned streaming guards instead classify partial output at token or sentence boundaries, trading timeliness against context, inference cost, and statistical uncertainty.

This paper asks a narrower engineering and conformance question. Suppose an application uses a fixed ordered policy of two-\allowbreak{}predicate conjunctive families. Can persistent state replace repeated whole-\allowbreak{}prefix scans while preserving the first withheld chunk under every finite chunking and in both production runtimes?

This question matters even when the lexical policy is deliberately small. Streaming APIs expose an irreversible sequence of release decisions, not one classification over a completed document. A monitor that eventually identifies the same pattern can still be operationally wrong if it releases the chunk that completed that pattern, blocks a different family when two complete together, or changes behavior when an SDK or upstream model partitions the same text differently. Conversely, a faster incremental recognizer is not a valid replacement merely because it finds the same matches on completed strings. Replacement requires equivalence at every release boundary.

The answer is not obtained by latching every temporary predicate match. A word boundary at the current end of a prefix is provisional: \texttt{\detokenize{\bcat\b}} matches \texttt{\detokenize{cat}}, but not the extended prefix \texttt{\detokenize{cats}}. Nor is a fixed overlap window enough for a same-\allowbreak{}line predicate containing an unbounded non-\allowbreak{}newline span. The implementation must carry recognizer state, distinguish provisional and stable acceptance, compose the predicates in policy order, and make the decision before forwarding the current chunk.

The paper contributes:

\begin{enumerate}
\item a declared ASCII policy grammar and cross-\allowbreak{}runtime semantics contract;
\item a predicate-\allowbreak{}state construction that preserves provisional end-\allowbreak{}boundary truth without unsound permanent predicate latching;
\item a grammar-\allowbreak{}scoped theorem equating the production incremental monitor with an absorbing cumulative oracle at every releasable boundary;
\item production-\allowbreak{}path Python and TypeScript integrations, a structurally realistic public surrogate, and a clean-\allowbreak{}revision differential artifact;
\item an empirical scaling comparison that retains all timing samples and keeps performance claims separate from conformance and safety effectiveness.
\end{enumerate}

We do not claim a new string-\allowbreak{}matching algorithm, a new general model of runtime enforcement, broad harm detection, or completeness of the deployed policy.

The evaluation is organized around three questions. RQ1 tests boundary-\allowbreak{}level equivalence to the buffered oracle. RQ2 tests whether the separately implemented Python and TypeScript paths implement one semantics rather than their host regex defaults. RQ3 measures scaling under a frozen, neutral, nonmatching profile. The formal claim answers RQ1 for the declared grammar; the finite campaigns test the implementations that instantiate it.

\section{Related work and novelty boundary}

\subsection{Incremental pattern recognition}

Aho-\allowbreak{}Corasick establishes efficient multi-\allowbreak{}pattern string matching with retained automaton state \cite{aho}. Thompson's construction supplies the classical route from regular expressions to NFA simulation \cite{thompson}. Production systems such as Hyperscan combine string and finite-\allowbreak{}automata components for high-\allowbreak{}throughput multi-\allowbreak{}pattern matching \cite{hyperscan}. These works mean that incremental literals, streaming regex recognition, and retained automaton state are prior art. Our delta is the policy-\allowbreak{}level composition needed for release equivalence: two-\allowbreak{}predicate conjunction, provisional end assertions, absorbing decisions, document-\allowbreak{}order priority, and a shared Python/TypeScript contract.

Regex implementation choice remains security-\allowbreak{}relevant. Extended constructs can produce denial-\allowbreak{}of-\allowbreak{}service behavior even in nonbacktracking systems \cite{redos}. We therefore use a deliberately small grammar rather than accepting either host runtime's general regex language, and we scope the theorem to that grammar.

\subsection{Runtime enforcement}

Security automata characterize execution-\allowbreak{}monitoring enforcement of safety policies \cite{schneider}. Edit automata extend the model with suppression, insertion, and transformation \cite{editautomata}. This work consequently does not claim that observing a trace and suppressing its next action is new. It instantiates that lineage at the LLM text-\allowbreak{}release boundary and establishes an equivalence result for the specific ordered-\allowbreak{}pair policy and its production integrations.

Stream runtime verification already separates declarative stream specifications from their online evaluation, storage, and update schedules \cite{booleansrv}. Prefix transducers incrementally match prefix expressions, retain configurations that may require more events, and permit explicit priority to resolve otherwise nondeterministic transitions \cite{prefixtransducers}. Those capabilities are also prior art. The narrower delta here is equivalence to a cumulative substring-\allowbreak{}search oracle whose temporary end-\allowbreak{}of-\allowbreak{}prefix assertions can change truth, composed with persistent two-\allowbreak{}predicate families and enforced before an LLM chunk is released.

\subsection{Streaming LLM safeguards}

SentGuard buffers sentence-\allowbreak{}sized units and releases classifier-\allowbreak{}verified chunks \cite{sentguard}. Qwen3Guard-\allowbreak{}Stream applies a token-\allowbreak{}level learned classification head for real-\allowbreak{}time moderation \cite{qwen3guard}. StreamGuard forecasts the expected harmfulness of likely continuations from partial prefixes \cite{streamguard}. SIREN (Safeguard with Internal REpresentatioN), the system introduced in "LLM Safety From Within," uses internal model representations for efficient streaming harm detection \cite{siren}. These systems seek broad semantic coverage and report statistical effectiveness. Our mechanism is model-\allowbreak{}agnostic, text-\allowbreak{}only, and exact only with respect to a narrow declared policy. It neither forecasts risk nor resolves paraphrase and context as a learned guard can.

Programmable application rails \cite{nemo} and systematic guardrail design \cite{systematic} further establish that runtime controls belong in layered, testable systems. This work contributes a precise conformance result inside that larger design space, not a complete safety architecture.

Table 1 states the novelty boundary in reviewer-\allowbreak{}facing terms.

\begin{table*}[t]
\caption{Novelty boundary relative to established capabilities}
\label{tab:1}
\centering\small
\setlength{\tabcolsep}{3pt}
\resizebox{\textwidth}{!}{%
\begin{tabular}{lll}
\toprule
\textbf{Prior capability} & \textbf{Already established} & \textbf{Question addressed here} \\
\midrule
Incremental literals and regular languages & Automata retain matching state across input & Do ordered two-\allowbreak{}predicate families preserve the same first withheld chunk? \\
Runtime suppression & Monitors may suppress an event or terminate a trace & What is the event and decision order at an LLM text-\allowbreak{}release boundary? \\
Stream runtime verification and prefix transducers & Online monitors retain state, consume prefixes, and may use priorities & How is a mutable end-\allowbreak{}of-\allowbreak{}prefix assertion made equivalent to cumulative substring search at every release boundary? \\
Learned streaming moderation & Partial outputs can be classified before completion & Can a narrow deterministic backstop provide exact, cross-\allowbreak{}runtime conformance? \\
High-\allowbreak{}throughput regex systems & Multiple patterns can be scanned efficiently & Can the production policy expose a small auditable grammar and reproducible oracle? \\
\bottomrule
\end{tabular}%
}
\end{table*}

This positioning leaves the strongest claim deliberately narrow. The result is not that automata can stream, but that this policy composition, assertion semantics, priority rule, and release integration are equivalent to the specified buffered behavior.

\section{Policy and release semantics}

\subsection{Threat and trust model}

The protected asset is text not yet released to the downstream client. The generator, its prompts, and the resulting output may be adversarially influenced. The adversary may choose any input text, induce arbitrary finite chunk partitions, split a match at every possible boundary, emit many distinct Unicode code points, and attempt to increase recognition cost or evade the configured signatures. The guard process, canonical policy artifact, release call order, and host runtimes are trusted. Compromise of those components, mutation of the policy during a stream, bypass of the guarded release path, and denial of service outside matcher state are excluded; Section 9 states the operational controls required to preserve those assumptions.

The security objective is deliberately narrow: before releasing chunk \texttt{\detokenize{c_j}}, the production monitor must return the same absorbing family decision as the declared cumulative oracle on prefix \texttt{\detokenize{x_j}}, independent of the finite chunk partition and host language. The objective includes bounded memoization state under adversarial text. It does not assert that the private signatures cover all harmful outputs, resist semantic paraphrase, or remain effective after policy disclosure.

The publication grammar accepts ASCII pattern source containing literals, character classes and ranges, ASCII-\allowbreak{}explicit \texttt{\detokenize{\s}}, \texttt{\detokenize{\w}}, and \texttt{\detokenize{\d}} categories and their negations, dot with LF excluded, \texttt{\detokenize{\b}}, grouping, alternation, concatenation, and bounded or unbounded quantifiers. ASCII simple folding maps only \texttt{\detokenize{A}} through \texttt{\detokenize{Z}}; word characters are exactly \texttt{\detokenize{[A-Za-z0-9_]}}; no Unicode normalization occurs. These choices eliminate accidental dependence on Python or ECMAScript regex behavior. Unicode homoglyph resistance is outside scope.

A policy is an ordered list of labelled families. Each family contains exactly two predicates. A family is currently satisfied when both predicates have matched somewhere in the accumulated prefix. If multiple families complete at one boundary, document order decides. After a block, the label is absorbing, because the completing chunk is withheld and the stream terminates from the client's perspective.

Formally, let \texttt{\detokenize{G}} be the grammar, \texttt{\detokenize{P = ((l_i, r_i1, r_i2))}} for \texttt{\detokenize{1 <= i <= m}} be an ordered policy, and \texttt{\detokenize{x_j = c_1 ... c_j}} be the prefix after chunk \texttt{\detokenize{j}}. Predicate truth is match-\allowbreak{}anywhere substring-\allowbreak{}search truth under the declared ASCII semantics: \texttt{\detokenize{search(r, x)}} holds exactly when there exist indices \texttt{\detokenize{a}} and \texttt{\detokenize{b}}, with \texttt{\detokenize{0 <= a <= b <= length(x)}}, such that an accepting run for \texttt{\detokenize{r}} spans \texttt{\detokenize{x[a:b]}}; assertions are evaluated against their actual predecessor and successor in \texttt{\detokenize{x}}, or against temporary start/end markers at the prefix boundary. A family is satisfied at \texttt{\detokenize{x_j}} exactly when both of its predicates have a match in \texttt{\detokenize{x_j}}. Let \texttt{\detokenize{first(P, x_j)}} be the smallest family index satisfying that condition, or bottom when none does.

The cumulative oracle appends a chunk, scans the full prefix under the declared semantics if no earlier block exists, chooses the first satisfied family, and then latches that decision. This is not a nonabsorbing fresh scan of later text: there is no later releasable boundary after a block.

The release contract is therefore:

\begin{scriptsize}
\begin{verbatim}
state := bottom
prefix := empty
for each candidate chunk c:
    if state is a label: return blocked(state)
    prefix := prefix || c
    state := first(P, prefix)
    if state is a label: withhold c; return blocked(state)
    release c
\end{verbatim}
\end{scriptsize}

The incremental monitor must produce the same \texttt{\detokenize{state}} before the \texttt{\detokenize{release c}} operation, without reconstructing or rescanning \texttt{\detokenize{prefix}}.

Figure 1 makes the irreversible order explicit: recognition, family priority, and the block decision all occur before the candidate chunk can be released.

\begin{figure*}[t]
\centering
\begin{tikzpicture}[
  node distance=7mm and 8mm,
  box/.style={draw, rounded corners=2pt, align=center, minimum height=9mm,
    text width=28mm, fill=black!3},
  decision/.style={draw, diamond, aspect=2.2, align=center, inner sep=1.5pt,
    fill=black!3},
  terminal/.style={draw, rounded corners=2pt, align=center, minimum height=9mm,
    text width=24mm},
  flow/.style={-{Latex[length=2mm]}, thick},
  verify/.style={-{Latex[length=2mm]}, dashed, semithick}
]
\node[box] (chunk) {candidate chunk $c_j$};
\node[box, right=of chunk] (recognizers) {persistent predicate\\recognizer states};
\node[box, right=of recognizers] (families) {ordered family\\conjunctions};
\node[decision, right=of families] (decision) {first label?};
\node[terminal, above right=3mm and 9mm of decision] (withhold) {withhold $c_j$\\latch label};
\node[terminal, below right=3mm and 9mm of decision] (release) {release $c_j$\\continue stream};
\node[box, below=13mm of families, text width=31mm] (oracle) {absorbing cumulative oracle\\verification mode only};
\draw[flow] (chunk) -- (recognizers);
\draw[flow] (recognizers) -- (families);
\draw[flow] (families) -- (decision);
\draw[flow] (decision) -- node[above left, font=\scriptsize]{label} (withhold);
\draw[flow] (decision) -- node[below left, font=\scriptsize]{$\bot$} (release);
\draw[verify] (chunk.south) |- (oracle.west);
\draw[verify] (oracle.east) -| node[pos=0.78, below, font=\scriptsize]{compare before release}
  (decision.south);
\node[draw, densely dotted, fit=(recognizers)(families)(decision), inner sep=4mm,
  label={[font=\scriptsize]above:release-boundary monitor}] {};
\end{tikzpicture}
\caption{Release-boundary composition. The candidate chunk is consumed into persistent
predicate states, ordered families are evaluated, and the decision is made before the
chunk can be released. A completed family withholds the same chunk that completed it.
The cumulative oracle is retained only for verification and rollback assurance}
\label{fig:release-boundary}
\end{figure*}
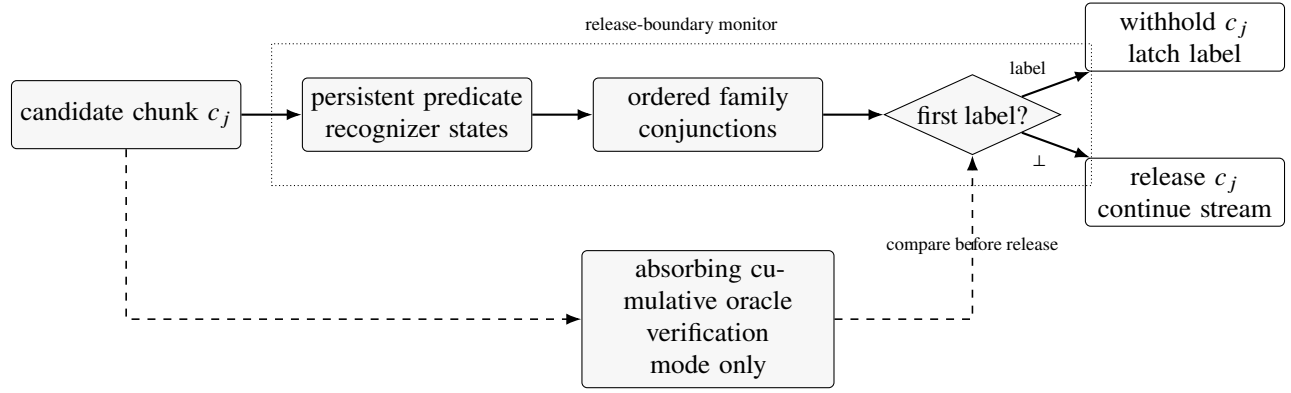

\section{Incremental construction and formal result}

Each predicate is compiled to a Thompson NFA. Its persistent state contains the active NFA states, the previous input character, and a stable-\allowbreak{}seen bit. Boundary transitions that need a following character are postponed. At a release boundary, the recognizer separately resolves those transitions against temporary end of input. A provisional predicate truth may disappear when the next character arrives; a stable match cannot.

\textbf{Worked surrogate trace.} Consider the public surrogate family whose predicates include \texttt{\detokenize{\bcolor\b}} and \texttt{\detokenize{\bshape\b}}. Table 2 feeds four candidate chunks to an initially empty stream. At boundary 1, \texttt{\detokenize{color}} matches only because temporary end of input supplies its trailing word boundary. The next chunk begins with \texttt{\detokenize{s}}, so that provisional acceptance disappears: \texttt{\detokenize{colors}} does not match \texttt{\detokenize{\bcolor\b}}. Boundary 3 presents \texttt{\detokenize{color }}; the space confirms the trailing word boundary, so the first predicate becomes stable. At boundary 4, \texttt{\detokenize{shape}} is true at temporary end of input and completes the family. The decision is made before release, so chunk 4 is withheld. A cumulative prefix scan makes the same four decisions.

\begin{table*}[t]
\caption{Worked public-surrogate trace showing provisional boundary acceptance, invalidation, stable acceptance, and pre-release withholding}
\label{tab:2}
\centering\small
\setlength{\tabcolsep}{3pt}
\resizebox{\textwidth}{!}{%
\begin{tabular}{llllll}
\toprule
\textbf{Boundary} & \textbf{Candidate chunk} & \textbf{Prefix at query} & \textbf{Color predicate} & \textbf{Shape predicate} & \textbf{Action} \\
\midrule
1 & \texttt{\detokenize{color}} & \texttt{\detokenize{color}} & true, provisional & false & release \\
2 & \texttt{\detokenize{s[SP]}} & \texttt{\detokenize{colors[SP]}} & false & false & release \\
3 & \texttt{\detokenize{color[SP]}} & \texttt{\detokenize{colors[SP]color[SP]}} & true, stable & false & release \\
4 & \texttt{\detokenize{shape}} & \texttt{\detokenize{colors[SP]color[SP]shape}} & true, stable & true, provisional & withhold chunk 4 \\
\bottomrule
\end{tabular}%
}
\end{table*}

Both monitors block at boundary 4 even though a continuation beginning with \texttt{\detokenize{s}} would falsify the \texttt{\detokenize{shape}} predicate. The declared semantics decide on the observed prefix, not on hypothetical extensions, and the absorbing oracle makes the same commitment.

The predicate-\allowbreak{}state lemma states that, after any consumed prefix, the retained state is the same NFA residual as a fresh substring search under the declared semantics, and \texttt{\detokenize{current()}} equals the fresh-\allowbreak{}search truth at that temporary end. The release-\allowbreak{}boundary theorem follows by induction over chunks: all predicate truths agree at a boundary, both monitors evaluate conjunctions in the same document order, and both decisions become absorbing at the first completion. Appendix A gives the definitions, induction invariant, lemma proof, and theorem proof within the manuscript; the tracked formal-\allowbreak{}claim package is a machine-\allowbreak{} auditable duplicate, not a substitute for peer review.

The temporary-\allowbreak{}end distinction is central. Before consuming the next character, the recognizer resolves pending boundary assertions against that character. After consuming it, the recognizer advances literal, class, and category edges and takes epsilon closure. At a release boundary only, a copy of the residual is resolved against the non-\allowbreak{}word end marker. Acceptance reached only through that temporary marker is reported by \texttt{\detokenize{current()}} but is not written to the stable-\allowbreak{}seen bit. Acceptance independent of the marker is stable and may be retained permanently.

\textbf{Lemma 1, predicate residual.} For every \texttt{\detokenize{r}} in \texttt{\detokenize{G}}, prefix \texttt{\detokenize{x}}, and chunking of \texttt{\detokenize{x}}, the incremental state after consuming \texttt{\detokenize{x}} represents the same residual language as a fresh NFA substring search, with following-\allowbreak{}character assertions pending. Its boundary query is true exactly when the fresh search is true on \texttt{\detokenize{x}} at temporary end of input.

\emph{Argument.} Induction over characters applies the same transition relation as the NFA, differing only in when assertions with an unknown successor are resolved. Seeding the start closure at every character implements substring search. The temporary end marker resolves the remaining pending assertions for the query without corrupting the residual retained for a possible extension.

\textbf{Theorem 1, release-\allowbreak{}boundary equivalence.} For every ordered pair policy over \texttt{\detokenize{G}}, every finite text, and every finite chunking, the incremental monitor and absorbing cumulative oracle return the same decision at every boundary up to and including the first block.

\emph{Argument.} Both begin at bottom. Assume their decisions agree before a chunk. If already blocked, absorption preserves equality. Otherwise Lemma 1 makes all predicate truths equal after the chunk; identical conjunction and document order therefore select the same family or bottom. The integration queries this decision before release, so the same completing chunk is withheld. Induction over boundaries completes the argument.

A bounded-\allowbreak{}overlap alternative is insufficient for the production grammar. For any overlap length, a same-\allowbreak{}line expression can place its opening literal, more than that many non-\allowbreak{}newline characters, and its closing literal in separate chunks. Persistent automaton state retains the required progress without retaining the whole line.

For fixed total automaton state count \texttt{\detokenize{Q}}, uncached incremental simulation uses \texttt{\detokenize{O(Q)}} active state and \texttt{\detokenize{O(nQ)}} transition work for \texttt{\detokenize{n}} input characters. The production implementation memoizes exact transitions from the active subset, predecessor wordness, and next input symbol to the next subset and acceptance bit, an on-\allowbreak{}demand DFA subset construction. The grammar distinguishes 128 ASCII symbols plus one equivalence symbol for every non-\allowbreak{}ASCII code point. One shared per-\allowbreak{}stream budget permits at most 4,096 cached transitions across all predicates; after the cap, unseen transitions are computed by the same NFA function but are not inserted. With cap \texttt{\detokenize{C}}, cache state is \texttt{\detokenize{O(CQ)}}. Repeated whole-\allowbreak{}prefix scanning performs work proportional to the sum of all prefix lengths and therefore grows quadratically under fixed-\allowbreak{}width chunking. The empirical section tests observed scaling separately.

\section{Production implementation}

The evaluated code is the matcher wired to Bee's Python and TypeScript customer-\allowbreak{}output paths, not a prototype, proof of concept, notebook rewrite, or paper-\allowbreak{}only fork. Both runtimes consume a generated view of one canonical policy artifact. The artifact records document-\allowbreak{}order priority and its SHA-\allowbreak{}256. The same generic matcher accepts the public surrogate; only the policy data changes.

The canonical JSON policy is the owned specification. Generation validates the grammar and emits the TypeScript view rather than maintaining two manually edited rule sets. Python instantiates the persistent monitor in the streaming safety path; TypeScript instantiates it in the customer-\allowbreak{}output path. Unit tests exercise parser rejection, assertions, classes, quantifiers, priority, and absorption. Integration tests exercise the pre-\allowbreak{}release call order. Cross-\allowbreak{}runtime verification exchanges only fixture text and decisions through a line-\allowbreak{}oriented process protocol.

Three runtime modes are retained. \texttt{\detokenize{incremental}} is the candidate production path. \texttt{\detokenize{cumulative}} is the reference behavior and is not used as evidence of speed. \texttt{\detokenize{verify}} runs both and fails closed on a decision mismatch while logging only policy identifier, label, boundary index, and mismatch metadata. It does not log generated content. An operator can select the prior cumulative mode as the rollback path while retaining the same policy artifact.

The production-\allowbreak{}path incremental recognizer memoizes NFA-\allowbreak{}subset transitions in both Python and TypeScript. The optimization is in the production classes, not a benchmark fork. A retained uncached option in the Python evidence harness permits exact differential and constant-\allowbreak{}factor measurement. The stream guard also records the prefix length already checked before release, so its terminal hook rescans only if a caller bypassed \texttt{\detokenize{feed}} and appended unchecked text.

The transition cache is per stream and shares a 4,096-\allowbreak{}entry budget across its predicate recognizers. Cache keys use the active subset, predecessor wordness, and one of 129 input symbols; every non-\allowbreak{}ASCII code point maps to the same symbol under the declared grammar. Once the budget is full, an unseen transition is computed but not cached. This prevents distinct-\allowbreak{}Unicode and unbounded-\allowbreak{}key cache inflation without changing decisions. Per-\allowbreak{}stream isolation prevents state crossing tenants, while aggregate memory still scales with active streams and must be controlled by deployment concurrency limits.

The release integration feeds a candidate chunk to the monitor before forwarding it. A non-\allowbreak{}null label withholds that chunk. Verification mode can compare the incremental decision with the cumulative oracle while recording only label and mismatch metadata, not generated content. Production rollout evidence remains distinct from implementation and conformance evidence.

\section{Evaluation}

\subsection{Questions}

RQ1 asks whether the production incremental matcher equals the absorbing cumulative oracle at every tested boundary. RQ2 asks whether Python and TypeScript implement the same declared semantics. RQ3 asks how incremental and cumulative modes scale in the frozen neutral-\allowbreak{}output profile and records the transition-\allowbreak{}cache state reached by adversarial fixtures.

\subsection{Frozen conformance campaign}

The campaign uses seed \texttt{\detokenize{260810279}}, 10,000 grammar-\allowbreak{}aware fixtures with exactly 10 randomized partitions each, and no-\allowbreak{}split plus every-\allowbreak{}character two-\allowbreak{}chunk coverage for each curated positive witness. Fixtures include exact matches, case changes, one-\allowbreak{}character near misses, newlines, ASCII boundary variations, and Unicode divergence probes. Every boundary compares Python incremental, Python cumulative-\allowbreak{}oracle, and TypeScript incremental decisions. Curated positives must also resolve to their intended family. The campaign includes a fixture containing 2,000 distinct non-\allowbreak{}ASCII code points and compares Python and TypeScript cache statistics as well as decisions.

The production policy has 37 curated witness combinations, producing 1,653 curated partition cases, plus 100,000 randomized partition cases. The public surrogate has 10 curated witnesses, producing 345 curated cases, plus 100,000 randomized cases. Any mismatch, crash, timeout, stale generated view, digest disagreement, missing response, or wrong intended family fails the run.

Random generation is grammar-\allowbreak{}aware rather than uniform character noise. For each construct it produces positives, case variants, one-\allowbreak{}character mutations, boundary substitutions, newline placements, start and end placements, and Unicode probes that must remain outside ASCII categories. Curated witnesses are split at every character boundary, preventing the randomized denominator from hiding a construct that never matches. Simultaneous-\allowbreak{}completion cases test document priority explicitly.

\subsection{Scaling and native-\allowbreak{}baseline profile}

Neutral nonmatching output is measured at 1,024, 2,048, 4,096, 8,192, and 16,384 characters with chunk sizes 8, 32, 64, 128, and 512. The comparators are the production memoized incremental monitor, the same Thompson-\allowbreak{}NFA monitor with memoization disabled, and cumulative full-\allowbreak{}prefix scanning by Python's native \texttt{\detokenize{re}} engine under \texttt{\detokenize{re.ASCII | re.IGNORECASE}}. The stability campaign specified five warmups and 25 retained repetitions per cell; no outlier is removed. The release-\allowbreak{}boundary feed path is timed without a redundant terminal rescan. The artifact retains every nanosecond sample, median, minimum, maximum, log-\allowbreak{}log least-\allowbreak{}squares slope, and both ratio directions. The earlier cumulative Thompson-\allowbreak{}NFA evidence remains a semantics/scaling comparator, but it is not presented as the practical native baseline. These measurements describe one machine and runtime; they are not universal latency estimates. All timed comparators in this subsection are Python implementations; the TypeScript path is tested for semantic and cache-\allowbreak{}stat conformance but is not used to claim cross-\allowbreak{}runtime performance parity.

The frozen machine was arm64 macOS 26.6 with Python 3.14.2, Node 26.4.0, and pnpm 11.21.0. The evidence runner first checked generated policy consistency, then ran Python tests, TypeScript typecheck and tests, the production and surrogate differential campaigns, and finally the benchmark. It recorded each command, return code, duration, environment key, input digest, Git status, and raw stdout and stderr. The conformance bundle used a clean worktree. The timing stability campaign ran from immutable commit \texttt{\seqsplit{e80a1efaf1159b20d14285b8fb7cd2833caeb556}}, addressed by annotated tag \texttt{\detokenize{p2-v2-evidence-1}}, in a detached temporary worktree and recorded clean status before and after execution. Neither run reported a failure or excluded sample.

\section{Results}

The definitive author-\allowbreak{}run bundle records clean Git revision \texttt{\seqsplit{e80a1efaf1159b20d14285b8fb7cd2833caeb556}}, canonical policy SHA-\allowbreak{}256 \texttt{\seqsplit{a97b282aae61d0627cea35ee478f7e4f9958ba98864e6e518d4b089abe9aa07f}}, and public-\allowbreak{}surrogate SHA-\allowbreak{}256 \texttt{\seqsplit{0cf7b1183252ca493f92b3bcfe4bec6b26c9c6a8f0ed63aa7e6453f458d41a87}}.

Table 3 reports the complete boundary-\allowbreak{}conformance campaign.

\begin{table*}[t]
\caption{Boundary-conformance outcomes for the production and public policies}
\label{tab:3}
\centering\small
\setlength{\tabcolsep}{3pt}
\resizebox{\textwidth}{!}{%
\begin{tabular}{llllllll}
\toprule
\textbf{Policy} & \textbf{Curated witnesses} & \textbf{Curated partitions} & \textbf{Seeded partitions} & \textbf{Total cases} & \textbf{Oracle mismatch} & \textbf{Cross-\allowbreak{}runtime mismatch} & \textbf{Wrong family} \\
\midrule
Production & 37 & 1,653 & 100,000 & 101,653 & 0 & 0 & 0 \\
Public surrogate & 10 & 345 & 100,000 & 100,345 & 0 & 0 & 0 \\
\bottomrule
\end{tabular}%
}
\end{table*}

All 101,653 production cases produced zero oracle, cross-\allowbreak{}runtime, or wrong-\allowbreak{}family mismatches. All 100,345 public surrogate cases also produced zero mismatches in those categories. These finite campaigns support implementation conformance; the universal claim depends on the grammar-\allowbreak{}scoped argument.

At 64-\allowbreak{}character chunks, the fitted log-\allowbreak{}log slopes were 0.972875 for the memoized incremental monitor, 1.002806 for the uncached incremental monitor, and 1.975511 for native cumulative regex scanning. At 16,384 characters their medians were 30.210, 216.480, and 96.557 ms, respectively. Transition memoization therefore reduced the incremental median by 7.17 times for this fixture and operating point.

Absolute performance depends on chunk size. At 16,384 characters the memoized incremental path was faster than native cumulative regex by 20.38, 6.13, 3.20, and 1.63 times at chunk sizes 8, 32, 64, and 128. At 512-\allowbreak{}character chunks the ordering reversed: native cumulative regex took 12.392 ms and incremental took 29.358 ms, so incremental was 2.37 times slower. Figure 2 and Table 4 expose this crossover directly. The clean run's 127.14-\allowbreak{}times cumulative-\allowbreak{}NFA comparison is retained only as implementation-\allowbreak{}internal scaling evidence; it is not used as a native-\allowbreak{}regex speed claim.

\begin{figure*}[t]
\centering
\begin{tikzpicture}
\begin{groupplot}[
  group style={group size=2 by 1, horizontal sep=1.45cm},
  width=0.465\textwidth,
  height=5.3cm,
  tick label style={font=\scriptsize},
  label style={font=\small},
  title style={font=\small\bfseries},
  grid=major,
  major grid style={black!12}
]
\nextgroupplot[
  title={(a) Size scaling at 64-character chunks},
  xmode=log, log basis x=2,
  ymode=log, log basis y=10,
  xlabel={Neutral input (characters)},
  ylabel={Median runtime (ms)},
  xtick={1024,2048,4096,8192,16384},
  xticklabels={{1,024},{2,048},{4,096},{8,192},{16,384}},
  legend style={font=\scriptsize, at={(0.02,0.98)}, anchor=north west,
    draw=none, fill=white, cells={anchor=west}}
]
\addplot+[color={rgb,255:red,0;green,114;blue,178}, mark=*, thick]
  coordinates {(1024,2.045834) (2048,3.924291) (4096,7.713416) (8192,15.270625) (16384,30.209625)};
\addlegendentry{Memoized incremental}
\addplot+[color={rgb,255:red,213;green,94;blue,0}, mark=square*, thick, dashed]
  coordinates {(1024,0.406875) (2048,1.542709) (4096,6.193500) (8192,24.239166) (16384,96.557167)};
\addlegendentry{Native regex cumulative}
\addplot+[color={rgb,255:red,0;green,158;blue,115}, mark=triangle*, thick, dotted]
  coordinates {(1024,13.511250) (2048,26.935625) (4096,53.518167) (8192,109.554375) (16384,216.480167)};
\addlegendentry{Uncached incremental}

\nextgroupplot[
  title={(b) Chunk-size crossover at 16,384 characters},
  xmode=log, log basis x=2,
  ymode=log, log basis y=10,
  xlabel={Chunk size (characters)},
  ylabel={Median runtime (ms)},
  xtick={8,32,64,128,512},
  xticklabels={8,32,64,128,512},
  legend style={font=\scriptsize, at={(0.02,0.98)}, anchor=north west,
    draw=none, fill=white, cells={anchor=west}}
]
\addplot+[color={rgb,255:red,0;green,114;blue,178}, mark=*, thick]
  coordinates {(8,37.756000) (32,31.430833) (64,30.209625) (128,29.640417) (512,29.357708)};
\addlegendentry{Memoized incremental}
\addplot+[color={rgb,255:red,213;green,94;blue,0}, mark=square*, thick, dashed]
  coordinates {(8,769.531583) (32,192.721542) (64,96.557167) (128,48.421542) (512,12.391667)};
\addlegendentry{Native regex cumulative}
\end{groupplot}
\end{tikzpicture}
\caption{Frozen timing-stability profile. Both panels report the median of 25 retained
repetitions after five warmups; no outlier was removed. Panel (a) separates linear
incremental scaling from native cumulative rescanning. Panel (b) exposes the absolute
constant-factor crossover: memoized incremental is faster through 128-character chunks,
while native regex is faster at 512. The result describes one machine and runtime}
\label{fig:scaling}
\end{figure*}
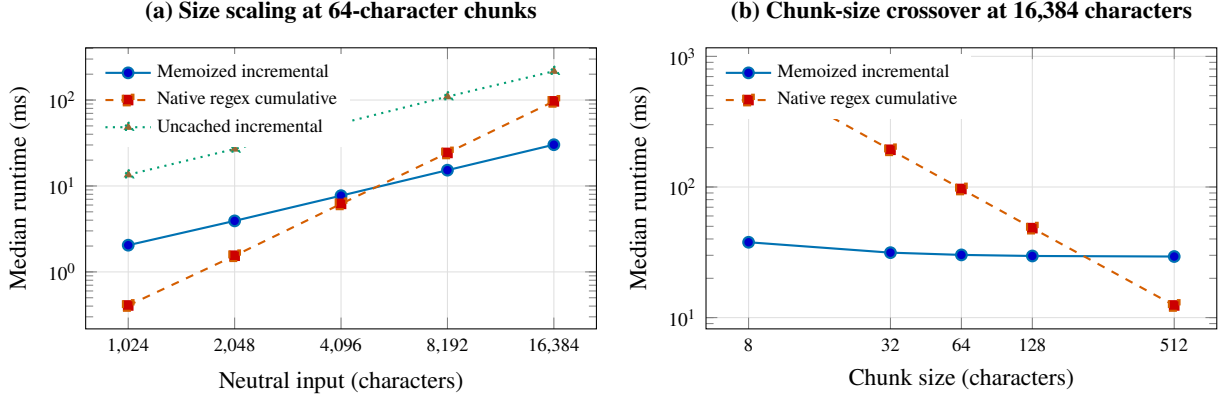

\begin{table*}[t]
\caption{Native-baseline crossover at 16,384 characters; each cell reports the median and observed range}
\label{tab:4}
\centering\small
\setlength{\tabcolsep}{3pt}
\resizebox{\textwidth}{!}{%
\begin{tabular}{lllll}
\toprule
\textbf{Chunk characters} & \textbf{Memoized incremental median ms [min, max]} & \textbf{Uncached incremental median ms [min, max]} & \textbf{Native cumulative regex median ms [min, max]} & \textbf{Native / memoized} \\
\midrule
8 & 37.756 [37.315, 38.376] & 227.873 [224.874, 249.903] & 769.532 [766.346, 776.807] & 20.38 \\
32 & 31.431 [31.022, 33.911] & 216.911 [214.263, 221.100] & 192.722 [191.303, 194.344] & 6.13 \\
64 & 30.210 [29.753, 31.358] & 216.480 [213.475, 222.018] & 96.557 [95.774, 97.215] & 3.20 \\
128 & 29.640 [29.305, 29.966] & 213.150 [209.196, 217.342] & 48.422 [47.931, 48.837] & 1.63 \\
512 & 29.358 [28.681, 30.501] & 209.040 [208.192, 212.158] & 12.392 [12.157, 12.641] & 0.42 \\
\bottomrule
\end{tabular}%
}
\end{table*}

Across the complete production campaign, the shared transition cache peaked at 364 entries (8.9\% of the 4,096-\allowbreak{}entry cap); the public surrogate peaked at 248 (6.1\%). The 16,384-\allowbreak{}character neutral profile used 123 entries at every tested chunk size. Neither campaign nor benchmark reached the cap, and Python and TypeScript reported identical cache statistics for every partition. One-\allowbreak{}entry regression tests exercise the overflow path and confirm that bypass preserves the uncached decision.

The native engine's favorable constant factor makes coarse cumulative scanning practical, exactly as the original study reported. The incremental path instead makes matcher work substantially insensitive to flush count and prevents quadratic growth when upstream systems emit fine chunks. Which path is faster at a bounded response length is therefore an operating-\allowbreak{}point question, not a consequence of the asymptotic result.

The bundle passed the frozen pre-\allowbreak{}evidence criteria and is admissible as author-\allowbreak{}run evidence. It is not an independent reproduction.

\section{Safety scope}

Conformance does not measure harm recall. The four production families are a small deterministic backstop, not a semantic moderation taxonomy. This work therefore does not reuse the original study's corpus or Llama Guard measurements as evidence for the new formal and implementation claim. It makes no contextual-\allowbreak{}disposition, false-\allowbreak{}positive, or broad harmful-\allowbreak{}output recall claim.

This separation prevents three common category errors. Zero conformance mismatches do not estimate false negatives against harmful language. The public surrogate validates mechanism reproducibility, not the quality of the private policy. A learned guard's broader classifications do not invalidate a deterministic backstop's exact-\allowbreak{}match function, just as the backstop's exactness does not establish semantic breadth. A deployed system may therefore layer these controls while evaluating each against its own claim.

\section{Operational implications}

The principal operational benefit is not simply lower asymptotic work. The stateful construction gives one decision contract to all chunk sizes and both runtimes. That makes upstream tokenization and network coalescing irrelevant to the configured family decision. The policy artifact can be reviewed separately from generated runtime views, and verification mode permits observation of candidate-\allowbreak{}versus-\allowbreak{}oracle disagreement without retaining customer text.

The same design creates obligations. Matcher state must be per stream and must not cross tenant or request boundaries. Policy updates cannot silently mutate an in-\allowbreak{}flight monitor; deployments must pin a policy version for the lifetime of each stream. A failover between runtimes must either transfer compatible state or restart from a retained prefix under the same policy. These lifecycle properties are integration requirements, not consequences of the theorem. The author-\allowbreak{}run operational-\allowbreak{}state evidence register names tests for independent matcher objects, immutable per-\allowbreak{}instance policy construction and digest exposure, and restart from a retained prefix under the same policy. Direct cross-\allowbreak{}runtime state serialization is not implemented or claimed; current failover semantics require retained-\allowbreak{}prefix replay.

\section{Limitations and threats to validity}

The policy covers only committed families and can miss paraphrases, reordered logic, obfuscation, code assembled at runtime, Unicode homoglyphs, fullwidth forms, or text outside its signatures. ASCII semantics are intentional for cross-\allowbreak{}runtime compositionality and auditability, not a claim of Unicode attack coverage. No normalization occurs, so normalization-\allowbreak{}based evasion is in scope as a limitation.

A safe prefix already released cannot be recalled. The guarantee is only that the first chunk making a configured family observable is withheld. A disclosed policy can invite adaptive evasion. The public surrogate establishes reproducibility of mechanism and grammar coverage, not disclosure or external validation of operational rules.

The differential campaign is finite and generator-\allowbreak{}dependent. Curated witness coverage reduces the risk of vacuous no-\allowbreak{}match testing but cannot enumerate all texts. The benchmark uses one machine, runtime versions, neutral input shape, and one policy. The work is authored and initially evaluated within one company. No independent clean-\allowbreak{}environment reproduction or non-\allowbreak{}author technical evaluation was available for the author-\allowbreak{}run results reported here.

The formal scope is also intentionally smaller than the parser's possible future feature set. Any grammar extension, normalization step, locale-\allowbreak{}aware fold, lookaround, backreference, or change to dot/newline behavior reopens the proof and cross-\allowbreak{}runtime gate. A passing campaign for the old grammar cannot be carried forward as evidence for a new one.

The cache cap bounds entries rather than bytes, and the subset stored in each entry remains bounded by the fixed policy automata. Its allocation is per stream, so aggregate cache memory grows linearly with concurrent streams even though one stream cannot exceed 4,096 entries. Deployment concurrency and maximum response length therefore remain operational resource controls. The campaign observed substantial headroom but does not enumerate every reachable subset of every future policy.

\section{Reproducibility, integrity, and version relationship}

The artifact provides a one-\allowbreak{}command clean-\allowbreak{}worktree verifier, canonical policy hashes, public surrogate, raw command logs, environment versions, raw timing samples, generated summaries, and a machine-\allowbreak{}readable manifest. An independent reproducer receives the frozen instructions and returns the evidence directory unchanged.

\textbf{Data and artifact availability.} The frozen public artifact, comprising the verifier, policy grammar, public surrogate, generators, tests, and expected author-\allowbreak{}run values, is deposited at Zenodo under doi:10.5281/zenodo.21932438. The deposit is cryptographically signed (ML-\allowbreak{}DSA-\allowbreak{}65) and its manifest records SHA-\allowbreak{}256 digests for every component. The original preprint is arXiv:2608.10279v1. No customer prompts, production traces, or private customer data are used in the public evaluation; third-\allowbreak{}party corpora used in internal review are identified by name and primary source and are not redistributed.

Every reference in this work is recorded in a machine-\allowbreak{}readable citation register with a primary authoritative URL, identifier, checked date, support relation, and disposition. The final citation keys must equal the registered included keys and the bibliography is hash-\allowbreak{}bound. Product identity and AI-\allowbreak{}assistance disclosure are not used as technical references.

arXiv:2608.10279v1 is the immutable original preprint corresponding to the prescreened journal submission. This work is a substantial post-\allowbreak{}decision revision with a new claim architecture, implementation, evidence campaign, public surrogate, and related-\allowbreak{}work analysis. This manuscript explicitly discloses that relationship and does not represent this revision as the version originally screened.

The public artifact exposes the matcher, grammar, surrogate policy, generators, tests, verifier, raw result schemas, and expected author-\allowbreak{}run values. The private review package binds operational-\allowbreak{}policy evidence by digest without placing its rules in third-\allowbreak{}party services. The reproduction protocol requires unedited output and a record of every undocumented intervention. No reproduction is called independent until a non-\allowbreak{}author completes and attests it.

\section{Conclusion}

Persistent automata are not new, and runtime suppression is not new. The result is the exact production composition required to make a narrow ordered-\allowbreak{}pair policy chunk-\allowbreak{}invariant at release time. Under the declared grammar, the incremental Python and TypeScript monitors agree with an absorbing cumulative oracle while avoiding repeated whole-\allowbreak{}prefix scanning. The guarantee ends at policy conformance. Semantic coverage, adversarial robustness, independent reproduction, and contextual judgment require separate evidence.

\appendix
\section{Formal definitions and proof}

Let \emph{word(z)} be true exactly for ASCII letters, digits, and underscore, and false for the start marker, end marker, every other ASCII character, and every non-\allowbreak{}ASCII code point. A boundary edge between predecessor \emph{p} and successor \emph{q} is enabled exactly when word truth differs between \emph{p} and \emph{q}. Literal comparison applies ASCII simple folding; character-\allowbreak{}class, category, and dot semantics are those declared in Section 3.

For a compiled predicate NFA \emph{N = (Q, s, f, E)}, define \emph{close(A, p, q, known)} as the least set containing \emph{A} and every state reachable by epsilon edges; it also follows boundary edges when \emph{known} is true and their boundary condition holds for \emph{p, q}. When \emph{known} is false, boundary edges remain pending. A substring search seeds \emph{s} at every possible start position. This includes the current end position, so predicates capable of accepting the empty string have their standard search meaning.

After consuming prefix \emph{x}, a predicate monitor stores the triple \emph{(A\_x, p\_x, h\_x)}. \emph{A\_x} is the active state set after the last consumed character and epsilon closure, with successor-\allowbreak{}dependent boundary edges pending; \emph{p\_x} is the final character of \emph{x}, or the start marker for empty \emph{x}; and \emph{h\_x} records an accepting run resolved independently of the temporary end marker. To consume character \emph{q}, the monitor adds start state \emph{s} to \emph{A\_x}, applies the closure with predecessor \emph{p\_x}, successor \emph{q}, and known assertions, follows every matching consuming edge, and applies closure again with unknown successor. It sets \emph{h} if acceptance is reached before or after the consuming transition. The boundary query returns true when \emph{h} is true or when closing \emph{A\_x} plus start state \emph{s} against the end marker reaches \emph{f}.

\textbf{Lemma A1, exact predicate query.} After every finite prefix \emph{x}, the stored triple has the definition above, and the boundary query is true exactly when \emph{search(r, x)} is true.

\emph{Proof.} Induct on the number of consumed characters. For the empty prefix, the stored active set contains no partially consumed run, the predecessor is the start marker, and the query seeds \texttt{\detokenize{s}} at position zero and closes it against the end marker. This is exactly fresh search on the empty string. Assume the invariant for \emph{x} and let \emph{q} be the next character. Every NFA run that can consume \emph{q} either began before the end of \emph{x}, in which case its state is in \emph{A\_x}, or begins at the new search position, represented by the added \emph{s}. A pending boundary now has its actual successor \emph{q}, so the first closure enables exactly the boundary edges enabled by the NFA semantics. Following all matching consuming edges and then epsilon-\allowbreak{}closing with successor unknown therefore produces exactly \emph{A\_(xq)}. Acceptance reached without using the temporary end marker is stable under extension and is recorded in \texttt{\detokenize{h_(xq)}}; no acceptance that depends only on the temporary marker is recorded there. Finally, the boundary query closes a copy against the end marker, thereby adding exactly the accepting runs valid only at the current end. Thus it equals fresh match-\allowbreak{}anywhere search on \emph{xq}. The invariant and equality follow by induction. Chunking only groups these character transitions and cannot change the resulting triple. End proof.

For policy \emph{P}, define \emph{F(P, x)} as the least document index whose two predicate queries are true on \emph{x}, or bottom if no family is satisfied. For chunks \emph{c\_1, ..., c\_t}, write \emph{x\_j = c\_1 ... c\_j}. The absorbing oracle state is \emph{B\_0 = bottom}; if \emph{B\_(j-\allowbreak{}1)} is a label then \emph{B\_j = B\_(j-\allowbreak{}1)}, otherwise \emph{B\_j = F(P, x\_j)}. The incremental monitor applies the same absorption rule to the family queries produced by its predicate states.

\textbf{Theorem A2, release-\allowbreak{}boundary equivalence.} For every policy over the declared grammar, text, and finite chunking, the incremental state equals \emph{B\_j} at every boundary through the first non-\allowbreak{}bottom state, and both monitors withhold the same first completing chunk.

\emph{Proof.} Induct on \emph{j}. Both initial states are bottom. If the common previous state is a label, absorption preserves equality. Otherwise Lemma A1 makes each incremental predicate query equal to fresh substring search on \emph{x\_j}. Consequently every two-\allowbreak{}predicate conjunction has the same truth value in both monitors. Both choose the least satisfied document index, so both select \emph{F(P, x\_j)}. The production integration queries this state before forwarding \emph{c\_j}; therefore the first non-\allowbreak{}bottom decision withholds that same chunk in both executions. End proof.

\textbf{Corollary A3, bounded memoized-\allowbreak{}transition equivalence.} Replacing a computed character transition by a cached value keyed by the complete active subset, predecessor wordness, and next grammar symbol preserves Lemma A1 and Theorem A2. The declared grammar maps all non-\allowbreak{}ASCII input to one behaviorally equivalent symbol, so the transition alphabet has 129 members. The key contains every input on which the transition function depends, and the cached value is the exact function result. When the shared per-\allowbreak{}stream cache reaches its cap, computing an unseen transition without insertion invokes that same function; cache hits, misses, overflow bypass, or eviction can therefore change cost but not the stored predicate state or release decision.

For completeness, no fixed suffix overlap implements the full declared grammar. Given overlap \texttt{\detokenize{k}}, place the opening literal of a same-\allowbreak{}line expression, more than \texttt{\detokenize{k}} non-\allowbreak{}newline characters, and its closing literal in later chunks. A suffix-\allowbreak{}only scanner loses the opening progress; the finite-\allowbreak{}state residual retains it. This statement distinguishes the construction from bounded-\allowbreak{}window rescanning without claiming that streaming regular-\allowbreak{}expression recognition is itself new.

\balance
\section*{Acknowledgements}
This research was conducted within HEOSSI (Pte.) Ltd.'s streaming-\allowbreak{}output assurance programme. BEE by HEOSSI (the Progressive Quantum-\allowbreak{}Native Intelligence Engine) supported research engineering, implementation analysis, evidence orchestration, and artifact verification under author supervision. Bee's production Python and TypeScript streaming paths provided the implementation context evaluated in this study. No customer material, production credentials, or private operational signature text is disclosed.

\section*{Statements and Declarations}
\noindent\textbf{Funding.} No external funds, grants, or other support were received for conducting this study or preparing the manuscript. The work used HEOSSI's internal research infrastructure and the support described in the Acknowledgements.

\noindent\textbf{Competing interests.} The author is Founder and Chief Technology Officer of HEOSSI (Pte.) Ltd., which develops Bee, whose streaming-\allowbreak{}output mechanism is evaluated in this study. HEOSSI may benefit from publication of results concerning its system. The author declares this employment, management, and commercial interest.

\noindent\textbf{Data and code availability.} The product-\allowbreak{}neutral artifact published at version DOI 10.5281/zenodo.21932438 (concept DOI 10.5281/zenodo.21932437) contains the declared grammar, public surrogate policy, generators, tests, verifier, result schemas, and author-\allowbreak{}run evidence. It excludes private operational signatures, customer material, and production credentials. The independent-\allowbreak{}reproduction record remains mandatory before external submission.

\noindent\textbf{Ethics.} The study used no human participants, personal data, customer prompts or responses, credentials, or live customer traffic. It inspected production integration code but did not experiment on customer-\allowbreak{}facing production systems. Ethics approval and consent were not applicable.

\noindent\textbf{Author contributions.} Christopher M. Frost conceived and supervised the study; defined its public, private, and operational boundaries; performed the methodology, software, validation, formal analysis, investigation, data curation, visualization, and writing; verified the claims, references, and evidence; and is responsible for the manuscript and release decision.

\noindent\textbf{Use of generative AI.} Generative AI tools provided general assistance with research organization, source review and manuscript drafting and editing.

\begingroup
\fontsize{7.5}{8.5}\selectfont

\endgroup
\end{document}